\documentclass[reprint,prl,superscriptaddress]{revtex4-2}

\usepackage{graphicx}
\usepackage{dcolumn}
\usepackage{bm}
\usepackage{hyperref}
\usepackage{amsmath}
\usepackage{hyperref}
\hypersetup{
	breaklinks=true,
	colorlinks=true,
	allcolors=blue
}

\begin{document}

\title{Monoenergetic acceleration of charge-neutralized ion bunches to GeV-scale energies\\ by the combination of a high-current electron beam and an ionization front}

\author{J. Chen}
 \affiliation{School of Applied and Engineering Physics, Cornell University, Ithaca, New York 14850, USA }

\author{J. Kim}%
\affiliation{%
School of Applied and Engineering Physics, Cornell University, Ithaca, New York 14850, USA
}%

\author{R. S. Rajawat}
\affiliation{School of Applied and Engineering Physics, Cornell University, Ithaca, New York 14850, USA
}%

\author{G. Shvets}
\affiliation{School of Applied and Engineering Physics, Cornell University, Ithaca, New York 14850, USA
}%
\affiliation{%
Cornell Laboratory for Plasma Studies (LPS), Cornell University, Ithaca, NY-14850, USA
}%

\date{\today}

\begin{abstract}
Compact heavy ion accelerators have numerous applications, ranging from heavy ion fusion to carbon ion radiotherapy, and testing radiation-hardened electronics. The demand could be met by developing high-gradient traveling wave plasma accelerators of high-charge ($\sim $ $\mu$C) relativistic ion beams. We will discuss a novel ion acceleration regime -- Counter-propagating ionization Front Acceleration (CFA) -- utilizing counter-propagating Ionization Front (IF) and high-current Relativistic Electron Beam (REB). Theoretical modeling and 3D PIC simulations demonstrate the possibility of using typical REBs produced by induction voltage adders propagating through a gas-filled tube undergoing laser ionization to achieve acceleration gradients in excess of $\sim 250 {\rm MeV/m}$ while accelerating micro-Coulombs of ions over meters distance. A unique energy conversion mechanism – from the REB to electromagnetic fields to the ions – is discussed, as well as the limits on the accelerated ions charge and the degree of its neutralization, acceleration gradient, and ion energy spread.

\end{abstract}

\maketitle


 High-intensity ion beam with GeV-scale energies are of great interest for a wide range of  applications, including heavy ion fusion\cite{meier_osiris_1992,waganer_inertial_1992}, carbon ion radiotherapy~\cite{solovyov_physics_2009,schardt_heavy-ion_2010}, fissile material production in spallation breeders ~\cite{fraser_high_1977}, and testing radiation-hardened electronics~\cite{schwank_radiation_2013}. Such ion beam accelerators typically use linear or circular metallic structures\cite{humphries_principles_2013}. However, their relatively low acceleration gradients (on the order of $1 {\rm MeV/m}$~\cite{seidl_p_a_notitle_2021} for linacs) and magnetic field limits (for cyclotrons)  result in a large physical footprint -- especially for heavy ions. Moreover, charge neutralization is required if ions need to propagate long distances to their final target\cite{henestroza_design_2004,kaganovich_physics_2010}.

Recent interest in plasma-based ion accelerators \cite{schreiber_review16,robinson_radiation_2008,schlegel_relativistic_2009,wilks_energetic_2001,ziegler_laser-driven_2024,park_ion_2019} arises from two attractive properties of the plasma: (i) its ability to sustain extremely high acceleration gradients and (ii) high degree of ion charge neutralization by the ambient plasma electrons.  For example, the most widely studied mechanism of ion acceleration in plasmas, Target Normal Sheath Acceleration (TNSA)\cite{wilks_energetic_2001,ziegler_laser-driven_2024},  produces accelerating fields on the order of a TV/m\cite{roth_ion_2016} at the rear surface of a solid target.  The TNSA scheme can be viewed as a form of {\it collective acceleration}: ions are pulled by the sheath electric field $E^{\rm (sh)}$ produced by a high flux of hot electrons generated at the front surface of the target by the incident laser pulse. The ion energy gain is limited by the product of the high $E^{\rm sh} \sim {\rm TV/m}$ and relatively short (microns) sheath length ($L^{\rm (sh)}$) . Therefore, the total energy of the accelerated ion beam is limited by the small acceleration volume defined by product of the acceleration length $L^{\rm (sh)}$ and its area constrained by the laser focal spot. In this Letter, we demonstrate how this constraint can be overcome by the combination of the two externally-supplied tools: (i) a large-area relativistic electron beam generated by an external accelerator (e.g., a high-current linac \cite{humphries_principles_2013}) and (ii) a sheath region moving synchronously with the accelerated ions, thereby trapping and accelerating them over much longer distances. In what follows, we show when compared with TNSA, the enhancement of the acceleration distance $L^{\rm (acc)} \gg L^{\rm (sh)}$ provided by (ii) more than compensates for the much weaker acceleration field  $E^{\rm sh} \sim 100 {\rm MV/m}$ and can produce GeV-scale ion bunches containing on the order of $10^{13}$ ions.

\begin{figure}[h]
\includegraphics[width=\columnwidth]{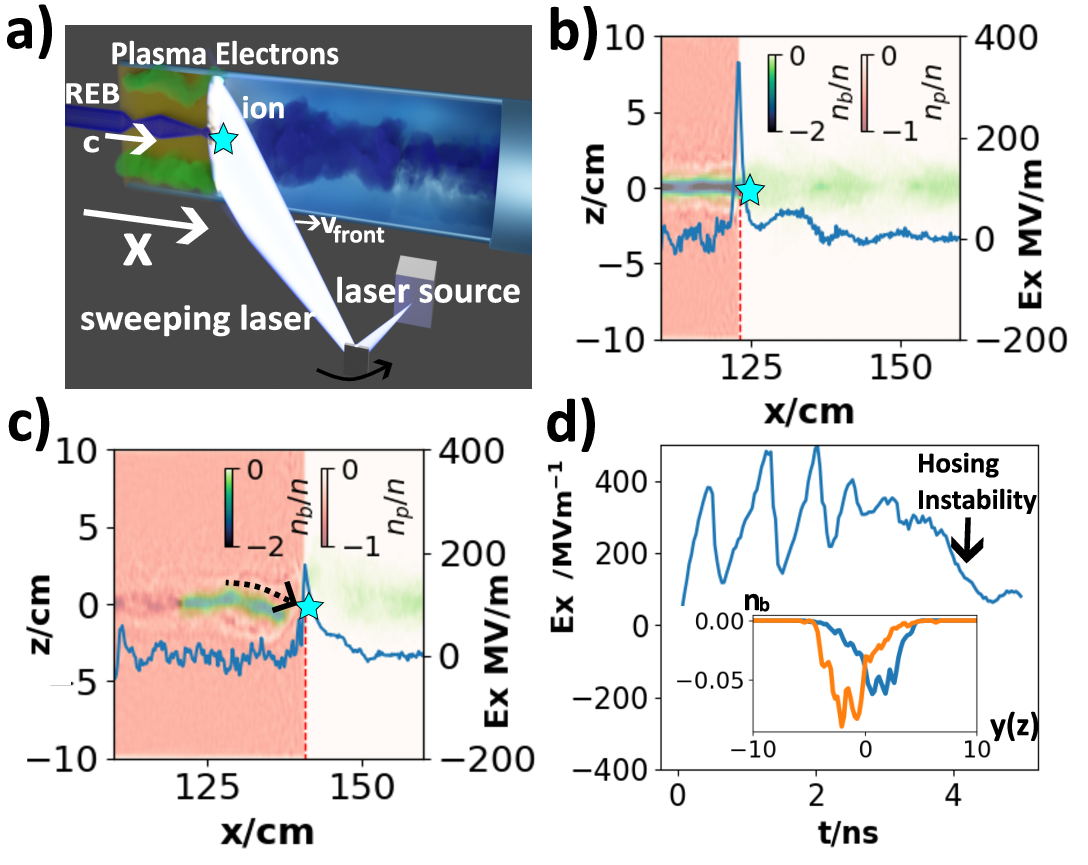}
\caption{\label{fig:fig1} (a) Schematic of an IFA: REB (blue), laser beam (white), and ions (cyan star) co-propagate from left to right. Plasma electrons/ions are shown as green/brown. (b-c) Time snapshots at $t_1 = 2.7 \mathrm{ns}$ and $t_2 = 4.1\mathrm{ns}$, respectively. Ionization front (vertical red dashed lines) moves from $x=100{\rm cm}$ at $t_0=0$ with the initial speed $v_{\rm IFA} = 0.2 c$. Plasma electron/REB densities (units: $10^{19} \,\mathrm{cm}^{-3}$) are color-coded in red/green. Blue curve: on-axis longitudinal electric field $ E_x^{\rm IFA}(x, t)$. (d)The evolution of $ E_{\rm x}^{\rm IFA}(x=x_{\rm IF}, t)$. Inset: REB density line-outs in $y/z$ (yellow/blue) at $x=x_{\rm IF}(t=t_2)$. }
\end{figure}

These two tools were first brought together in a collective ionization front accelerator (IFA) schematically shown in Fig.~\ref{fig:fig1}(a). ~\cite{olson_PoF75,olson_experimental_1986,olson_collective_1979,oshea_apl86}. In an IFA, an intense relativistic electron beam (REB; shown as the dark blue region) is directed through the column of gas towards the moving laser-generated interface between neutral and partially-ionized gas. The ionization front (IF) interface is produced by a side-sweeping laser beam (marked by the white color) ionizing the gas. The IF speed $v_{\rm IF} \equiv dx_{\rm IF}/dt$ is controlled by that of the laser sweep if the gas density is sufficiently low to avoid impact ionization by the REB~\cite{olson_pra75} while its molecular composition (e.g., alkaline metal vapor) is selected to enable ionization at modest laser intensities~\cite{olson_experimental_1986}. The ions co-propagating with the IF and the REB with the speed $v_{\rm i}(t)$ can continuously gain energy from the longitudinal accelerating field $E^{\rm IFA}$ inside the IF if their speeds are matched over time: $v_{\rm i}(t) = v_{\rm IF}$.

While the electric field of a relativistic charge is primarily transverse to its velocity, the mechanism of generating the longitudinal sheath field can be understood by observing that the ionized plasma electrons are repelled by the transverse electric force of the REB, leaving behind a positively charged ion channel. This establishes a sheath field $E^{\rm IFA}_{x}$ pointing from the downstream (ionized, positively charged) toward the upstream (neutral gas) regions separated by the IF, with its trajectory $x_{\rm IF}$ determined via the ionizing laser sweep controlled by either an electro-optical deflector\cite{romer_electro-optic_2014} or a flying focus\cite{pigeon_ultrabroadband_2024,palastro_ionization_2018}. The resulting accelerating gradient generated by a high-current (tens of kA) REB produced by a typical linear induction accelerator~\cite{smith_prstab04} can reach $ > 200\,\mathrm{MV/m}$, i.e. at least $2$ orders of magnitude higher than in conventional linear ion accelerators. Compared with other plasma-based ion accelerators\cite{gong_laser_2024}, IFA can accelerate very large charges $\sim\mathcal{O}(\mu \mathrm{C})$ owing to macroscopic spot-sizes of REBs  $(\sim\mathcal{O} (\rm cm))$. Ion acceleration over tens of centimeters in an IFA has been experimentally demonstrated ~\cite{olson_experimental_1986,oshea_apl86}.

\begin{table*}[t]
	\caption{\label{tab:simulation-parameters}%
		Simulation parameters used for the IFA and CIFA cases.}
	\begin{ruledtabular}
		\begin{tabular}{lcccc}
			Parameter & IFA & CIFA-1 & CIFA-2 & CIFA-3 \\
			\colrule
			REB duration, $\tau$ (ns)
			& 1.33 & 4.6 & 4.6 & 4.6 \\
			
			REB cutoff time, $t_{\mathrm{cutoff}}$ (ns)
			& 5 & 9.67 & 9.67 & 9.67 \\
			
			REB size parameter, $\omega$ (cm)
			& 0.707 & 0.707 & 0.707 & 0.707 \\
			
			Initial REB kinetic energy, $U_{k0}^{\mathrm{REB}}$ (MeV)
			& 10 & 10 & 2 & 2 \\
			
			Initial REB density, $n_{b0}$ ($\mathrm{cm}^{-3}$)
			& $5\times10^{12}$ & $5\times10^{12}$
			& $5\times10^{12}$ & $5\times10^{12}$ \\
			
			Initial gas electron density, $n_{e0}$ ($\mathrm{cm}^{-3}$)
			& $5\times10^{12}$ & $5\times10^{12}$
			& $5\times10^{12}$ & $5\times10^{12}$ \\
			
			Accelerated-ion charge, $Q_a$
			& test ion & test ion
			& $0$--$250~\mathrm{nC}$
			& $0$--$5.4~\mu\mathrm{C}$ \\
			
			Front parameter, $\alpha$
			& 0.7 & 0.7 & 0.7 & 0.3 \\
		\end{tabular}
	\end{ruledtabular}
\end{table*}

Below we use first-principles particle-in-cell (PIC) simulation code \textsc{Smilei}~\cite{derouillat_smilei_2018} to demonstrate that extending the IFA length is challenging because of beam-plasma instabilities, such as the electron hose instability of high-current REBs propagating through ion channels  ~\cite{whittum_prl91,lampe_electronhose_1993} previously identified in {\it un-magnetized} plasmas. To overcome the free expansion of REB in the vacuum \cite{Stupakov_2021_classical} due to its self-field and mitigate the violent pinching inside the plasma, an axial magnetic field $ \mathbf{B}_{\mathrm{ext}} = 1\,\mathrm{T}\, (\hat{\mathbf{x})} $ is applied. We further assume equal peak REB and gaseous plasma densities: $n_{\rm b0} = n_{\rm g0} = 5 \times 10^{12} \,\mathrm{cm}^{-3}$, and the REB kinetic energy of $U^{\rm REB}_{\rm k0} = 10 \,\mathrm{MeV}$. Detailed simulation parameter is shown inside the Table~\ref{tab:simulation-parameters}. The spatio-temporal REB profile is assumed in the following form:
\begin{equation}\label{eq:profile}
  n_{\rm b} = n_{\rm b0}\exp\!\left[-\frac{(t - t_{0})^{2}}{2\tau^{2}} - \frac{r^{2}}{2w^{2}}\right]
\end{equation}
for $|t - t_{0}| < t_{\mathrm{cut}}/2$, where $\tau = 1.33\,\mathrm{ns}$, $t_{\mathrm{cutoff}} = 5\,\mathrm{ns}$, and $w = 0.7\,\mathrm{cm}$. In Fig.~\ref{fig:fig1} the IF sweep starts at $x_0=100\mathrm{cm}$ and proceeds in the positive $x$-direction with a initial velocity $ v_{\rm IF}(t=0) = 0.2c$. The comparison of the snapshots of the peak on-axis acceleration fields shown in Figs.~ \ref{fig:fig1}(b–c) clearly shows that the magnitude of the accelerating field $E_{\rm x}^{\rm IFA}$ rapidly decreases between $t_1 =2.7 {\rm ns}$ and $t_2 = 4.1 {\rm ns}$ .

This reduction is further illustrated in Fig.~\ref{fig:fig1}(d), where the peak $E_{\rm x}^{\rm IFA}(x=x_{\rm IF},t)$ at the IF location $x_{\rm IF}(t)$ is plotted as a function of the propagation time (corresponding to the total propagation distance of $x_2 - x_0 \approx 40\mathrm{cm}$). We note that a strong oscillatory behavior of $E_{\rm x}^{\rm IFA}$ for earlier times $t < t_2$ corresponds to the REB over-focusing because of the mismatch between its initially small size/emittance and the strong focusing strength of the ion channel~\cite{noauthor_linear_2008}. Our simulations indicate that the troughs of the accelerating field cooresponds to the times when the IF encounters the focusing position of the REB. These pinching-generated oscillations are reduced for longer propagation time as the REB emittance increases.

While it may be possible to avoid REB pinching by the appropriate choice of the initial REB emittance, the eventual reduction of the on-axis accelerating field cannot be overcome because of the severe electron-hose instability~\cite{whittum_prl91,lampe_electronhose_1993}: the beam centroid at the IF is displaced in both transverse directions as shown in the inset in Fig. \ref{fig:fig1}(d). To the best of our knowledge, neither the electron-hose instability in a {\it magnetized} plasma channel, nor its limiting effect on the peak ion energies produced by IFAs have been previously investigated.

\begin{figure}[htbp]
\includegraphics[width=\columnwidth]{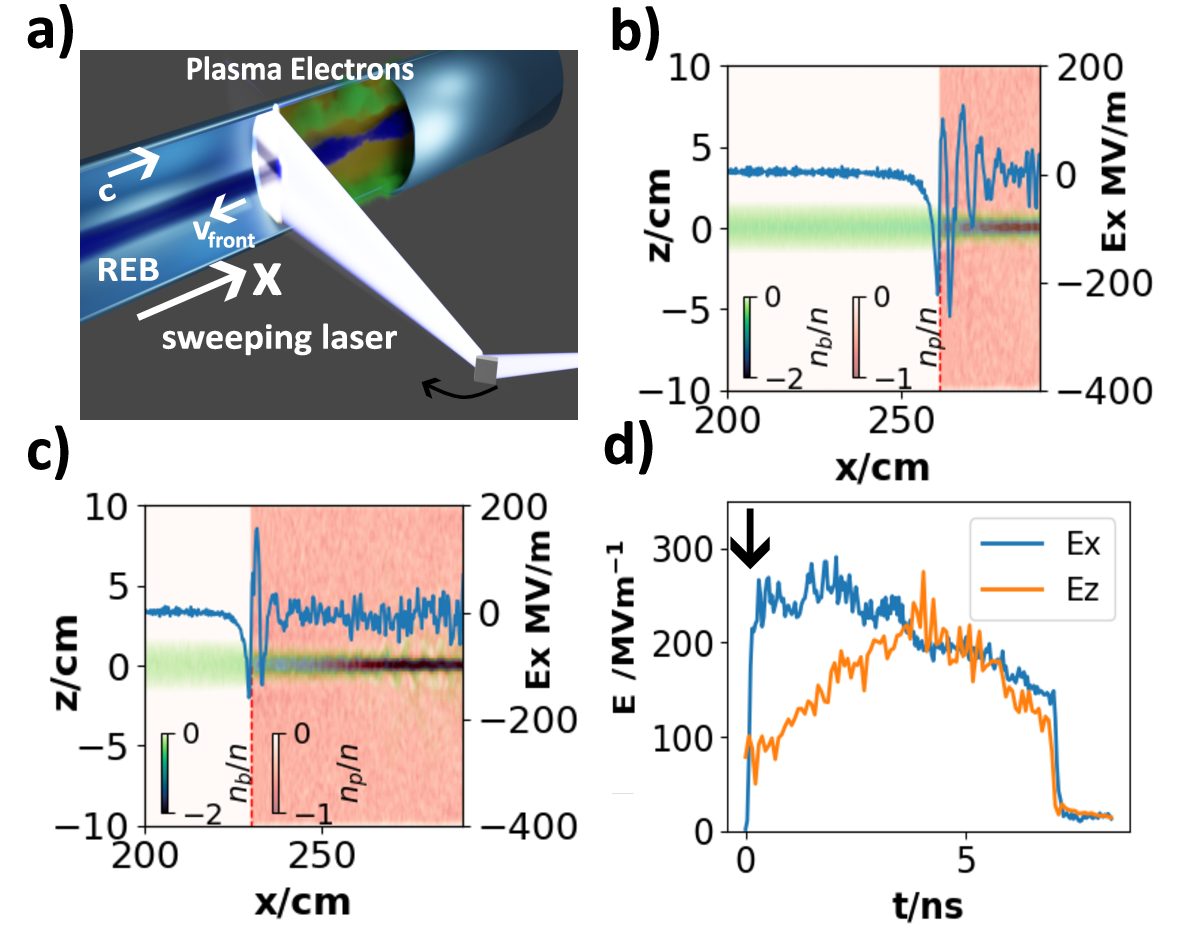}
\caption{\label{fig:fig2} (a) Schematic of a counter-propagating IFA (CIFA). Coloring scheme: same as in IFA in Fig.~\ref{fig:fig1}(a), but the REB and IF/ions counter-propagate in CIFA.  (b-c) Time snapshots at $t_1=3.7\mathrm{ns}$ and $t_2=6\mathrm{ns}$, respectively.  Ionization front (vertical red dashed lines) moves from $x_0 = 295{\rm cm}$ at $t_0=0$ with the speed $v_{\rm CIFA} = 0.2 c$. Plasma electron/REB densities (units: $10^{19} \,\mathrm{cm}^{-3}$) are color-coded in red/green. Blue curve: on-axis longitudinal electric field $ E_x^{\rm CIFA}(x, t)$. (d)The evolution of the peak value of the on-axis longitudinal (blue line: $E_x$)  and the transverse focusing (yellow line: $E_z$) electric field components near the location of the IF. The arrow denotes the time of encounter of REB and IF.}
\end{figure}

To entirely avoid the electron-hose instability and restrict the pinching of the REB, we propose a new concept, termed as \textit{counter-propagating ionization front acceleration} (CIFA), where the external laser sweeps in the direction opposite to the REB propagation. All other parts of IFA are kept same as illustrated in Fig.\ref{fig:fig2}(a). Due to the counter-propagating configuration, the IF continuously encounters a fresh flow of the REB that has not yet interacted with the plasma. Consequently, the accelerating field at the front remains unaffected by the pinching or beam-plasma instability of the REB.

To validate the CIFA concept, we performed a series of PIC simulations with REB and plasma parameters identical to those used in Fig.  \ref{fig:fig1} and the following characteristics of the counter-propagating IF (CIF),also shown in colomn CIFA-1 in Table~\ref{tab:simulation-parameters}:  the IF starts at $x_{\rm CIF} (t=0) = 295\mathrm{cm}$ and propagates with initial velocity of $v_{\rm CIF}(t=0) = -0.2c$. The REB profile is given by Eq.(\ref{eq:profile}), where $\tau = 4.6\,\mathrm{ns}$, and $t_{\mathrm{cutoff}} = 9.67\,\mathrm{ns}$ and the REB first encounters the CF at $t=0$ at the $x_0 = 295{\rm cm}$ location.
As shown in Figs.~\ref{fig:fig2}(b–d), the accelerating field at the counter-propagating IF maintains nearly-constant value of $E_{\rm x}^{\rm CIFA}(x=x_{\rm CIF},t) \sim250\mathrm{MV/m}$ and exhibits no temporal oscillations or decrease due to REB hosing -- in stark contrast to the co-propagating IFA case depicted in Fig.~\ref{fig:fig1}(d). In contrast to the limited acceleration distance $L^{\rm IFA}_{\rm acc}\approx 40 \mathrm{cm}$ in IFA, CIFA enables meter-scale acceleration distance $L^{\rm CIFA}_{\rm acc}\approx 0.85 \mathrm{m}$. Consequentially for developing GeV-scale ion accelerators is that the acceleration distance can be further extended using longer REB durations, as indicated by the sharp cutoff of the electric field at $t=7\mathrm{ns}$ corresponding to the time instance of the REB tail colliding with the IF. Additionally, the yellow curve in Fig. \ref{fig:fig2}(d) highlights the strong transversely-focusing electric field generated by the REB, which confines the accelerated ions near the beam centroid. The resulting focusing of the ions operates similarly to the Gabor lens \cite{gabor_nature47,gabor_fnal_90}, but with an important difference: both the radial electric and azimuthal magnetic fields of the REB contribute to the strong focusing of high-charge ion bunches and prevent their space charge-driven expansion.

In the rest of this Letter, we (i) discuss the peak amplitude $E_{\rm acc}$ of the accelerating field $E_{\rm x}^{\rm CIFA}$ near the IF, and (ii) investigate the limits on the ion bunch charge $Q_i$ trapped by the IF that is swept forward with a uniform acceleration $g$ according to 
$d \left( \gamma_f \dot{x}_f(t) \right) = g $, where $\gamma_f = 1/\sqrt{1 - \beta_f^{2}}$ and $\beta_f = \dot{x}_f/c$. 

In either co- or counter-propagation geometry, $E_{\rm acc}$ originates in the charge separation between plasma electrons and ions, with the electrons being pushed out by the REB to the charge-neutralization radius~\cite{whittum_prl91} $a=\sqrt{n_b/n_i} r_b$, where $n_b$ and $r_b$ are the density and radius of the REB, respectively, and $n_i$ is the ion density. This expression assumes uniform, top-hat transverse density profiles. 

Employing the quasi-static approximation (QSA) in the vicinity of the ionization front, we introduce the co-moving coordinate
$\xi=x-v_{\rm CIF}t$ and the pseudo-potential
$\psi=\Phi-v_{\rm CIF}A_x$. Then $E_x^{\rm CIFA}$ can be given by(see the Supplemental Material):
\begin{equation}\label{eq:field_solve}
	\begin{aligned}
		\left(
		\nabla_{\perp}^{2}
		+ \frac{1}{\gamma_{\rm CIF}^{2}}
		\frac{\partial^{2}}{\partial \xi^{2}}
		\right)\psi
		&=
		-\frac{1}{\epsilon_0}
		\left(
		\rho_e
		-\beta_{\rm CIF}\frac{J_x}{c}
		\right),
		\\[4pt]
		E_x^{\rm CIFA}
		&=
		-\frac{\partial\psi}{\partial\xi},
	\end{aligned}
\end{equation}
where $\beta_{\rm CIF}=v_{\rm CIF}/c$ and
$\gamma_{\rm CIF}=(1-\beta_{\rm CIF}^2)^{-1/2}$. Here, $\rho_e$ and
$J_x$ denote the net charge density and net longitudinal current density,
respectively.

We further assume a highly relativistic REB and calculate the axial electric
field produced by a uniformly charged cylinder of radius $a$ that is
semi-infinite along the $x$ direction and consists of immobile plasma ions.
The characteristic field amplitude at the end of this ion column is
\begin{equation}\label{eq:E0}
	E_0
	=
	\frac{e n_i a}{2\epsilon_0}
	=
	\frac{e\sqrt{n_b n_i}\,r_b}{2\epsilon_0}.
\end{equation}
Simulations show that a fraction of the newly ionized plasma electrons is
injected into the channel by the longitudinal accelerating field. The
density of these injected electrons depends on $v_{\rm CIF}(t)$. Motivated by the
$\gamma_{\rm CIF}^{-2}$ term in Eq.~\eqref{eq:field_solve}, and accounting
for the contribution of the injected electrons to the effective source,
$E_{acc}$ can be expressed as the ion-column
field multiplied by a monotonically decreasing correction factor
$0<f(\lvert v_{\rm CIF}\rvert)\leq1$ (see the Supplemental Material):
\begin{equation}\label{eq:peak_field}
	E_{acc}
	=
	f\!\left(\left|v_{\rm CIF}\right|\right)E_0
	=
	f\!\left(\left|v_{\rm CIF}\right|\right)
	\frac{e\sqrt{n_b n_i}\,r_b}{2\epsilon_0},
\end{equation}

The injection of plasma electrons is driven by the combined action of the strong axial electric field \(E_{\rm acc}\) and the azimuthal magnetic field \(B_\theta\), produced by the REB current. Immediately after being ionized, the plasma electrons are accelerated by \(E_{\rm acc}\) to a relativistic longitudinal velocity \(v_x\). Their longitudinal motion through \(B_\theta\) then produces a radial magnetic Lorentz force. This magnetic force counteracts the outward electric force, allowing the injected electrons to remain confined within the ion channel.

Phase stability is one of the necessary conditions for any accelerator because it ensures that the charged particles with slightly different initial conditions can undergo long-term acceleration~\cite{edwards_syphers}. The standard approach to investigating phase stability that we adopt in this Letter is to introduce a synchronous ion accelerated with a constant accelerating gradient $E_f \equiv \alpha E_{acc} < f E_0$ inside the ionization front moving with it, here we denote its newton acceleration $g = Z_i e E_f / M_i$, where $eZ_i$ and $M_i$ are the ion electric charge and mass, respectively. For simplicity, we assume $f = 1$, i.e. neglect its velocity dependence, which can be suppressed by designing the appropriate temporal density profile of the REB. We employ the Hamiltonian analysis to study the phase stability of an ion whose phase space position $\left( x,p_x \right)$ is close to that of the synchronous one, e.g., their relative separation $\xi = x - x_f$ is small. The relativistic equation of motion for the synchronous ion co-moving with the front is given by $d \left( \gamma_f \dot{x}_f(t) \right) = g $, where $\gamma_f = 1/\sqrt{1 - \beta_f^{2}}$ and $\beta_f = \dot{x}_f/c$. The analytic solution for $\dot{x}_f(t)$ is given by
\begin{equation}
x_f(t)
= \frac{c}{g}
\sqrt{\,c^{\,2} + \left( g t + \gamma_0 \beta_0\, c \right)^{2}}
\;-\;
\frac{c^{2}}{g}\gamma_0,
\label{eq:xf}
\end{equation}
where $\gamma_0 = 1/\sqrt{1 - \beta_0^{2}}$ and $\beta_0 = v_0/c$ are the initial front speed. 
The Hamiltonian is expressed as \cite{gong_laser_2024}:
\begin{equation}
H = \frac{1}{2} m_i\, \dot{\xi}^{\,2}
\;+\;
\frac{1}{\gamma_f^{3}(t)}
\left( m_i\, g\, \xi + e\,\Phi(\xi) \right)
= T + f(t)\,V(\xi) .
\end{equation}

Here, the electric field ($E(\xi)$) and the electrostatic potential ($\Phi(\xi)$) in this region are determined from the semi-infinite column charge distribution.

\begin{figure}[htp]
\includegraphics[width=\columnwidth]{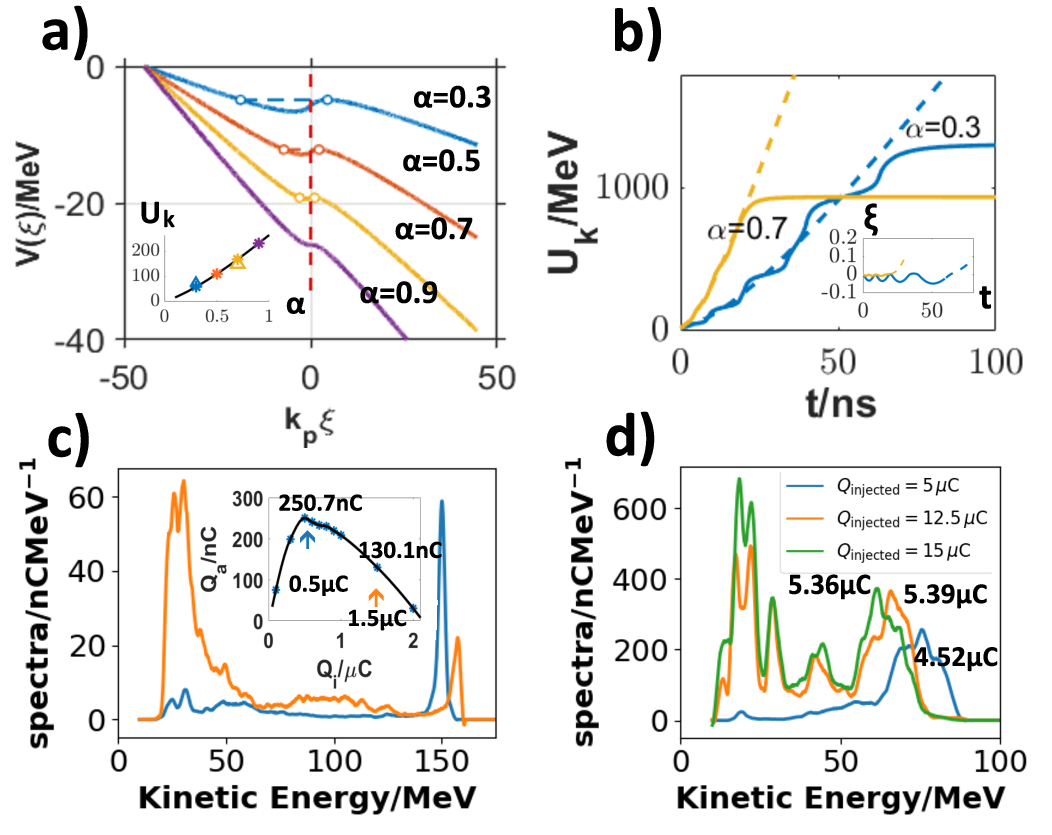}
\caption{\label{fig:fig4}
(a) Potential energy distribution $V(\xi)$ from Eq.~(6) for different $\alpha$. The inset shows the theoretical final kinetic energy $U_k/\mathrm{MeV}$ versus $\alpha$ (black curve); the star and triangle denote the theoretical and simulation values, respectively.
(b) Numerical solution of the Hamiltonian equations of motion. The main panel shows the evolution of $U_k(t)$ for the simulated ion (solid) and an imaginary ion co-moving with the IF (dashed). The inset shows $\xi(t)$, with solid and dashed segments indicating the motion before and after escape from the potential well. Colors denote different $\alpha$.
(c–d) Spectra of accelerated ions for $\alpha=0.7$ (c) and $\alpha=0.3$ (d). Different curves correspond to different injected charges $Q_i$. In (c), the blue and orange curves correspond to $Q_i=0.5\mu\mathrm{C}$ and $1.5\mu\mathrm{C}$, respectively; the inset shows the accelerated-peak charge $Q_a$ at $150\mathrm{MeV}$ versus $Q_i$, with arrows indicating the two curves in the main panel. In (d), $Q_i$ is given in the legend, and the accelerated peak is labeled next to each curve.}
\end{figure}

By varying the ratio $\alpha$, we obtain different effective potentials $V(\xi)=m_i\, g\, \xi + e\,\Phi(\xi)$, as shown in Fig.~\ref{fig:fig4}(a).
It is found that potential wells are only formed at $\alpha < 1$.
For slowly varying $\gamma_f$, since the relative speed starts at zero, this suggests ions get trapped, oscillate in the well, and undergo sustained acceleration.
The length of the potential well increases as $\alpha$ decreases, indicating a greater capacity to trap ions for smaller $\alpha$.
In contrast, when $\alpha > 1$, the potential no longer forms a well, and ion trapping—and thus acceleration—does not occur.

For a given beam duration $t_{\mathrm{cutoff}}$, the total ion acceleration time $T$ can be calculated using the relation $|x_f(T)| + cT = c\,t_{\mathrm{cutoff}}$, which describes the condition where the IF encounters the tail of the REB at time $T$. Assuming $\gamma_0\beta_0 << 1$, analytic solution of $T$ is $t_{\mathrm{cutoff}}(2c+gt_{\mathrm{cutoff}})/2(c+gt_{\mathrm{cutoff}}+c\gamma_0\beta_0)$.
The corresponding final kinetic energy is then given by $U_k = U_{k0} + E_f \left( c\,t_{\mathrm{cutoff}} - cT \right)$, where $U_{k0}$ is the ion's initial kinetic energy, assuming a constant $E_f$ at the front.
The inset in Fig.~\ref{fig:fig4}(a) shows a monotonic increase in the final ion energy with increasing $\alpha$, obtained for $t_{\mathrm{beam}} = 10\,\mathrm{ns}$, for the case of proton.
To conclude, the value of $\alpha$ can serve as a fine-tuning parameter to control the final kinetic energy and the maximum accelerated charge.

The time dependence of the Hamiltonian becomes increasingly important at larger $\gamma_f(t)$ and over longer acceleration times. In Fig.~\ref{fig:fig4}(b), we numerically solve the Hamiltonian equations for a proton initially at $\xi=0$ with zero velocity relative to the front. See the inset plot, lines are split into solid and dashed parts. Ion is initially trapped and oscillates in the potential well, as in the quasi-static case (solid inset line), but escapes as the sweeping velocity increases and $\gamma_f$ rises more rapidly (dashed inset line). The main panel shows the corresponding energy evolution. For a long acceleration time ($\sim100\mathrm{ns}$), the ions reach maximum energies of $\sim900\mathrm{MeV}$ for $\alpha=0.7$ and $\sim1300\mathrm{MeV}$ for $\alpha=0.3$ before de-trapping. The maximum energy also depends on the initial ion position: ions starting farther from the potential minimum escape earlier and gain less energy. Therefore, smaller $\alpha$ not only improves trapping, but also enables higher maximum ion energy for sufficiently long REB duration, whereas larger $\alpha$ yields faster energy gain over a shorter distance but a lower final energy.

Simulations are performed to study proton trapping and acceleration for different $\alpha$, with details found in CIFA-2 and CIFA-3 colomn of Table~\ref{tab:simulation-parameters}. To maximize energy-conversion efficiency, a $2\mathrm{MeV}$ REB is used, with a total acceleration time of $\sim7\mathrm{ns}$. Protons are injected with velocity matched to the CIF, $v=-0.2c$, and an emittance of $10\mathrm{mm\cdot mrad}$.
Figures~\ref{fig:fig4}(c,d) show the final ion spectra for $\alpha=0.7$ and $0.3$. For $\alpha=0.7$, quasi-monoenergetic acceleration is obtained at $\sim150\mathrm{MeV}$. As the injected charge increases, the monoenergetic peak charge first saturates and then decreases due to beam loading, reaching a maximum of $\sim250\mathrm{nC}$ [inset of Fig.~\ref{fig:fig4}(c)].
In the case of $\alpha = 0.3$, shown in Fig.~\ref{fig:fig4}(d), the peak energy is reduced to approximately $80\,\mathrm{MeV}$, and the energy emittance becomes greater.
Nevertheless, the charge of the accelerated ion peak exceeds $5\,\mu\mathrm{C}$—which is nearly two orders of magnitude more than typical TNSA\cite{hornung_enhancement_2020}, with a monoenergetic charge profile.
These accelerations correspond to the initial stage of Fig.~\ref{fig:fig4}(b), where the yellow curve is higher than the blue curve; the higher accelerated charge capacity for $\alpha = 0.3$ is also consistent with the potential distribution illustrated in Fig.~\ref{fig:fig4}(a). The corresponding energy-conversion efficiencies are $0.8\%$ for $\alpha=0.7$ and $6.5\%$ for $\alpha=0.3$. These values are obtained without optimization of the injected ion temporal profile, which is left for future study.

In conclusion, we have proposed and investigated a novel ion acceleration scheme—counter-propagating ionization front acceleration (CIFA).
Through particle-in-cell simulations, we demonstrate that this configuration circumvents the hosing instability and avoids the oscillatory electric fields inherent in traditional IFA.
Hamiltonian analysis reveals that ion trapping and maximum energy gain are strongly dependent on the field ratio $\alpha = E_f / E_{acc}$.
Simulations with moderate parameters show that, in less than one meter, CFA enables a smooth tuning from quasi-monoenergetic ion beams with energies exceeding $150\,\mathrm{MeV}$ and accelerated charge up to $250\,\mathrm{nC}$, to total accelerated charges above $5\,\mu\mathrm{C}$ with energies of $\sim75\,\mathrm{MeV}$—surpassing state-of-the-art plasma-based acceleration methods by nearly two orders of magnitude.
Such energy and charge can be further extended by increasing the REB radius, density, or duration to accommodate higher-energy applications.
These findings establish CFA as a modular, programmable, high-charge, and high-energy ion acceleration regime, opening pathways for compact, laser-controlled ion sources in high-energy physics, heavy-ion fusion, and medical applications.


\bibliographystyle{apsrev4-2}
\bibliography{CFApaper}

\end{document}